\documentclass[10pt]{article}
\usepackage{multicol}
\usepackage{geometry}
\usepackage{xcolor}

\definecolor{citecolor}{RGB}{31,120,30}
\definecolor{urlblue}{RGB}{30,30,200}
\usepackage[hidelinks,colorlinks,linkcolor=urlblue,citecolor=citecolor]{hyperref}

\usepackage{fontsize}
\usepackage{graphicx}
\usepackage{float}

\usepackage{amsmath}
\usepackage{mathtools}
\usepackage{gensymb}

\usepackage{booktabs}
\usepackage{multirow}
\usepackage{makecell}
\usepackage[default]{fontsetup}

\usepackage{microtype}
\usepackage{changepage}
\usepackage{physics2}
\usephysicsmodule{braket}
\usepackage{upgreek}
\usepackage[labelfont=bf]{caption}

\usepackage[
	backend=biber,
	style=numeric-comp,  
	sorting=none,        
	maxbibnames=6,       
	minbibnames=3,       
	doi=true,            
	url=true,            
	isbn=true,           
	giveninits=true,     
	terseinits=false     
]{biblatex}

\AtEveryBibitem{%
	\iffieldundef{doi}
		{}
		{\clearfield{url}%
		 \clearfield{urlyear}%
		 \clearfield{urlmonth}%
		 \clearfield{urlday}}%
}

\definecolor{authorcolor}{RGB}{0,0,0}

\renewcommand{\title}[1]{{\LARGE\bfseries\raggedright{#1}\par\vskip.15in}}
\renewcommand{\author}[1]{{\raggedright\large\textbf\boldmath\color{authorcolor}#1\vskip1ex\par}}
\newcommand{\address}[1]{{\raggedright\small #1\par}}
\newcommand\authormark[1]{\textsuperscript{#1}}
\newcommand{\email}[1]{{\raggedright\footnotesize\itshape\color{urlblue}{#1}}\par}
\renewenvironment{abstract}{\vskip1pc\noindent\textbf{Abstract:}}{}

\begin{document}
\begin{adjustwidth}{2cm}{2cm}
\title{Demonstration of a Monolithic and Fully Telecom-Fiber-Compatible Tunable Source of Polarization Entangled Photon Pairs Based on a van der Waals Material}

\author{Benjamin Laudert,\authormark{1,2*} Fatemeh Abtahi,\authormark{1} Duk Choi\authormark{3}, Dragomir Neshev\authormark{2}, and Falk Eilenberger\authormark{1,4,5}}

\address{
	\authormark{1}Institute of Applied Physics, Abbe Center of Photonics, Friedrich Schiller University Jena, Albert-Einstein-Str.~15, 07745 Jena, Germany\\[2pt]
	\authormark{2}ARC Centre of Excellence for Transformative Meta-Optical Systems (TMOS), Department of Electronic Materials Engineering, Research School of Physics, Australian National University, Canberra, ACT, Australia\\[2pt]
	\authormark{3}Department of Quantum Science and Technology, Research School of Physics, Australian National University, Canberra, ACT, Australia\\[2pt]
	\authormark{4}Fraunhofer-Institute for Applied Optics and Precision Engineering IOF, Albert-Einstein-Str.~7, 07745 Jena, Germany\\[2pt]
	\authormark{5}Max Planck School of Photonics, Albert-Einstein-Str.~15, 07745 Jena, Germany
}

\email{\authormark{*}benjamin.laudert@uni-jena.de}

\begin{abstract}
We present a tunable, single-mode-optical-fiber-based source of polarization entangled photon pairs for the near-infrared telecommunication band that is deployable in standard infrastructure. The photon pairs are generated via spontaneous parametric down-conversion (SPDC) in a submicron-scale thin film of the inversion-broken rhombohedral polytype of the transition metal dichalcogenide molybdenum disulfide (3R-MoS\textsubscript{2}), located between two fiber connectors. By exploiting the intrinsic symmetries of the second-order nonlinear susceptibility tensor of 3R-MoS\textsubscript{2}, this hybrid approach offers control over the generated two-photon polarization state through the incident pump polarization. Most notably, two of the four maximally entangled Bell states, as well as fully co-polarized pairs can be produced. This represents a substantial improvement in terms of tunability and simplicity over established fiber-integrated sources, which require additional optical elements, precise alignment, or careful engineering of design parameters.
Additionally, a nonlinear drop in the background photoluminescence signal of 3R-MoS\textsubscript{2} is observed at low pump powers, allowing us to reach a coincidences-to-accidentals ratio (CAR) of $(8.3\pm1.8)\times10^{3}$, the highest value recorded for SPDC in van der Waals materials to date.
\end{abstract}
\end{adjustwidth}

\vskip1cm
\begin{multicols}{2}
\section{Introduction}
Entangled photon pairs are a highly versatile tool for a wide variety of quantum applications, such as quantum key distribution (QKD) \cite{ekertQuantumCryptographyBased1991, mubashirkhanHighErrorrateQuantum2009}, entanglement swapping \cite{panExperimentalEntanglementSwapping1998}, quantum repeaters \cite{briegelQuantumRepeatersRole1998} and several quantum imaging and sensing techniques
\cite{
	pittmanOpticalImagingMeans1995,
	botoQuantumInterferometricOptical2000,
	abouraddyQuantumopticalCoherenceTomography2002,
	nagataBeatingStandardQuantum2007,
	lloydEnhancedSensitivityPhotodetection2008},
that offer capabilities beyond classical optics.
For many of these applications, entangled photon sources based on single-mode optical fibers are highly desirable, as they greatly simplify handling and deployment in real-world environments and enable direct integration with existing fiber infrastructure.
Furthermore, several applications benefit greatly from photon pairs that are entangled in the polarization degree of freedom, as they can be manipulated and analyzed with well-established polarization hardware, while some advanced QKD protocols \cite{mubashirkhanHighErrorrateQuantum2009} make use of many different multi-photon states, driving the need for tunable entangled photon sources.

A promising method for the generation of entangled photons, which underpins many state-of-the-art sources, is spontaneous parametric down-conversion (SPDC). In SPDC, a single pump photon is split into an entangled pair of signal and idler photons under conservation of energy and momentum \cite{harrisObservationTunableOptical1967}. SPDC is a second-order nonlinear process and requires a medium with a non-zero second-order nonlinear susceptibility tensor $\chi^{(2)}$, i.e.~broken inversion symmetry \cite{boydNonlinearOptics2020}.

Over the past decades, tremendous progress has been made in the highly efficient generation of photon pairs via SPDC.
However, most of the well-established platforms, such as birefringent \cite{kwiatNewHighIntensitySource1995, kwiatUltrabrightSourcePolarizationentangled1999} or periodically-poled crystals \cite{kuklewiczHighfluxSourcePolarizationentangled2004}, do not intrinsically generate polarization entangled photons and, thus, require additional post-selection, arrival-time difference compensation, or interferometry schemes \cite{kimPhasestableSourcePolarizationentangled2006}.
Furthermore, their integration with single-mode optical fibers is often cumbersome and requires precise alignment and mode-matching, custom-tailored to the fiber and the SPDC source spatial emission pattern. This presents a severe complication for the wide-scale deployment of entangled photon sources in existing long-range telecommunication infrastructure, which is mostly based on near-infrared single-mode fibers.

In monolithic devices, such as waveguides \cite{
	levinePolarizationentangledPhotonPairs2011,
	hornInherentPolarizationEntanglement2013,
	sunCompactPolarizationentangledPhotonpair2019,
	warkeDirectGenerationTwopair2022,
	zhangPolarizationentangledPhotonPairs2025},
photonic chips \cite{zhukovskyGenerationMaximallypolarizationentangledPhotons2012},
nano-structured materials \cite{
	maPolarizationEngineeringEntangled2023,
	jiaPolarizationentangledBellState2025,
	moPolarizationentangledPhotonPairs2025},
or periodically poled fibers
\cite{
	zhuDirectGenerationPolarizationEntangled2012,
	chenBroadbandFiberbasedEntangled2021a},
design parameters must be carefully engineered and maximal polarization entanglement is linked to wavelength-specific optical modes for the pump, signal, and idler radiation, which complicates tuning of the generated state without the use of additional optical elements.
Exploring compact, tunable, and robust methods for the generation of entangled photon pairs with direct integration into optical fibers, therefore, remains crucial for the scalable deployment of many quantum technologies.

In recent years, a variety of vdW materials with broken inversion symmetry have emerged as promising nonlinear media for SPDC \cite{
	guoUltrathinQuantumLight2023,
	weissflogTunableTransitionMetal2024,
	fengPolarizationentangledPhotonpairSource2024,
	guoPolarizationEntanglementEnabled2024,
	trovatelloQuasiphasematchedDownconversionPeriodically2025,
	liangTunablePolarizationEntangled2025,
	lyuTunableEntangledPhotonpair2025,
	luCounterpropagatingEntangledPhoton2025,
	linNonlinearPhasematchedVan2026,
	joshiInLineFiberIntegratedPhotonPair2026a}.
These materials consist of individual crystal sheets bound to each other by vdW forces, which allows for bulk crystals to be thinned down to films of arbitrary thickness through mechanical exfoliation These can be integrated into many well-established photonic platforms, including optical fibers \cite{linNonlinearPhasematchedVan2026, joshiInLineFiberIntegratedPhotonPair2026a}, through scalable, low-cost polymer-based dry-stamping techniques.
Compared to their conventional counterparts, these materials possess greater nonlinearities \cite{guoUltrathinQuantumLight2023, hongTwistPhaseMatching2023, fengPolarizationentangledPhotonpairSource2024, trovatelloQuasiphasematchedDownconversionPeriodically2025}, which could enable the fabrication of more compact entangled photon pair sources with superior spectral bandwidths.

\begin{figure}[H]
	\centering
	\includegraphics{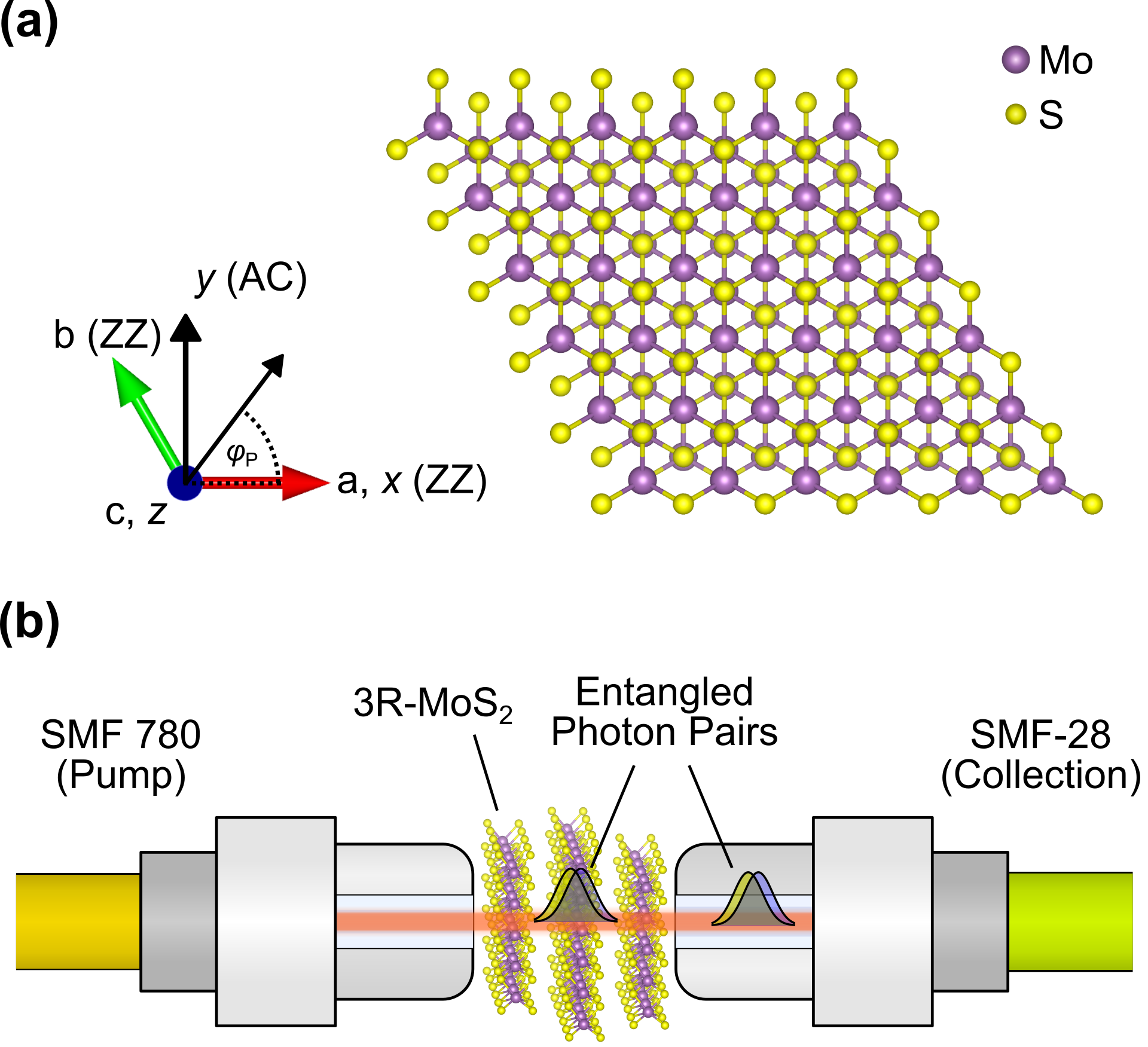}
	\caption{
		\textbf{In-line polarization-tunable entangled photon pair generation in 3R-MoS\textsubscript{2}.}
		\textbf{(a)} 3R-MoS\textsubscript{2} crystal structure as viewed along the c-axis (stacking direction) and coordinate system of the $\chi^{(2)}$ tensor elements, where $x$ and $y$ correspond to one of the zig-zag (ZZ) and armchair (AC) directions of the crystal, respectively. The pump polarization angle $\varphi_\mathrm{P}$ is defined relative to $x$.
		\textbf{(b)} Sketch of the entangled photon pair generation in the in-line geometry. The pump light is incident along the c-axis, allowing for tuning of the generated polarization state via the pump. 
	}
	\label{fig:crystal_structure}
\end{figure}

Furthermore, a subset of these materials, namely rhombohedral-phase molybdenum disulfide (3R-MoS\textsubscript{2}) \cite{weissflogTunableTransitionMetal2024}, tungsten disulfide (3R-WS\textsubscript{2}) \cite{fengPolarizationentangledPhotonpairSource2024}, and boron nitride (rBN) \cite{hongTwistPhaseMatching2023}, possess $C_{3v}$ point group symmetry.
Their in-plane $\chi^{(2)}$ tensor components, oriented perpendicular to their inter-layer stacking direction, are therefore highly symmetric \cite{boydNonlinearOptics2020}:
\begin{align}
	- \chi^{(2)}_{yyy} = \chi^{(2)}_{yxx} = \chi^{(2)}_{xxy} = \chi^{(2)}_{xyx} \label{eq:chi2_tensor}.
\end{align}
Here $x$ and $y$ correspond to the zig-zag (ZZ) and armchair (AC) directions of the crystal lattice as illustrated in Figure~\ref{fig:crystal_structure}(a) for the case of 3R-MoS\textsubscript{2}.
This $\chi^{(2)}$ tensor symmetry enables the generation of polarization entangled photon pairs with consistent polarization states over broad spectral windows, without carefully chosen device parameters or the need for any additional optical elements.

Moreover, the generated state can be tuned based on the incident pump
polarization state $\ket{\psi_\mathrm{P}}$.
Depending on the pump polarization angle $\varphi_\mathrm{P}$ and retardance $\delta_\mathrm{P}$, where
\begin{align}
	\ket{\psi_\mathrm{P}} = \cos(\varphi_\mathrm{P}) \ket{H} + e^{i\delta_\mathrm{P}} \sin(\varphi_\mathrm{P}) \ket{V} \label{eq:pump_state}
\end{align}
is defined as a superposition of horizontal $\ket{H}$ and vertical $\ket{V}$ polarization states, the generated two-photon polarization state is \cite{shindeGenerationTunableEntanglement2026}:
\begin{align}
	\begin{split}
		\ket{\psi_\mathrm{G}} &= \frac{\sin(\varphi_\mathrm{P})}{\sqrt{2}} \left( \ket{HH} - \ket{VV} \right)\\
		&\quad +\frac{e^{i\delta_\mathrm{P}} \cos(\varphi_\mathrm{P})}{\sqrt{2}} \left( \ket{HV} + \ket{VH} \right).
	\end{split} \label{eq:gen_biph_state}
\end{align}
Note that, here, we use a contracted notation for the Kronecker product of the signal (s) and idler (i) ket states, e.g.
$\ket{HV} = \ket{H}_\mathrm{s} \otimes \ket{V}_\mathrm{i}$.

When the pump state is circularly-polarized, i.e. $\varphi_\mathrm{P} = 45\degree$ and $\delta_\mathrm{P} = \pm 90\degree$, where $+$ and $-$ respectively correspond to right $\ket{R}$ and left $\ket{L}$ circular polarization, a fully co-polarized state with inverted helicity, i.e. $\ket{LL}$ or $\ket{RR}$, is generated \cite{fengPolarizationentangledPhotonpairSource2024}.
In the case of linear pump polarization ($\delta_\mathrm{P} = 0$), the pump state $\ket{\varphi_\mathrm{P}}$ is in a balanced superposition of $\ket{R}$ and $\ket{L}$ and the generated two-photon state is maximally entangled \cite{weissflogTunableTransitionMetal2024, fengPolarizationentangledPhotonpairSource2024}.
Specifically, for $\varphi_\mathrm{P} = 0\degree$, one can define the horizontal and vertical polarization states $\ket{H}$ and $\ket{V}$ as logical qubit states $\ket{0}$ and $\ket{1}$, respectively.
$\ket{\psi_\mathrm{G}}$ then represents one of the four Bell states, $\ket{\psi^+} = \frac{1}{\sqrt2} (\ket{01} + \ket{10})$, which is of great interest for many of the aforementioned applications.
Keeping this choice of basis fixed while advancing $\varphi_\mathrm{P}$ by 90° leads to the generation of a second Bell state $\ket{\phi^-} = \frac{1}{\sqrt2} (\ket{00} - \ket{11})$.

VdW materials of the $C_{3v}$ symmetry group, thus, enable the generation of photon pairs spanning the whole range of polarization entanglement that can be tuned rapidly based on the incident pump polarization.
Previous studies of fiber-integrated vdW materials have, however, neglected to exploit this through polarization control \cite{linNonlinearPhasematchedVan2026} or used a material with different symmetry properties that make it unable to produce maximal polarization entanglement \cite{joshiInLineFiberIntegratedPhotonPair2026a}.

In this work, we demonstrate, for the first time, that this method for the tunable generation of polarization entangled photon pairs is fundamentally compatible with fully in-line placement of the nonlinear crystal between two standard FC/PC fiber connectors, as illustrated in Figure \ref{fig:crystal_structure}(b). This enables direct integration into existing telecom infrastructure, and thus, represents a potentially game-changing approach for the deployment of entangled photon sources for long-range fiber-based QKD in real-world systems.

We show that, even without prior knowledge of the polarization transformations arising from stress-induced birefringence, which transform the polarization states of the pump and downconverted photons as they traverse their respective fibers, a self-consistent unitary description of the fiber network, based on a reference polarization state at the pump wavelength, can be found with a reasonable number of test inputs and a simplified quantum state tomography measurement of the photon pair as a feedback element.
This enables deterministic control of the two-photon polarization state at the fiber output by tuning of the reference state with a standard fiber polarization controller.

We verify the state tuning methodology with polarization-resolved coincidence measurements that accurately reproduce the results of equivalent free-space counterparts, which have previously been performed in References \cite{weissflogTunableTransitionMetal2024, liangTunablePolarizationEntangled2025, lyuTunableEntangledPhotonpair2025}.

We use 3R-MoS\textsubscript{2} as a nonlinear medium. The material offers a large second-order nonlinear susceptibility of $\chi^{(2)} \approx 100\,\mathrm{pm}\mathrm{V}^{-1}$ \cite{trovatelloQuasiphasematchedDownconversionPeriodically2025}, combined with a sufficiently large 1.85 eV optical bandgap \cite{xuCompactPhasematchedWaveguided2022}, to generate photon pairs inside the low-loss telecom C-band, with minimal absorption of the pump radiation. Compared to previous studies of fiber-integrated vdW crystals \cite{linNonlinearPhasematchedVan2026, joshiInLineFiberIntegratedPhotonPair2026a}, where photon pairs are generated at a wavelength of $\lambda \approx 800\,\mathrm{nm}$, outside of the near-infrared telecom bands, this represents a substantial step towards practical deployment in existing infrastructure.

\section{Principle and Results}\label{sec:results}
\subsection{Setup and Sample}

\begin{figure*}[htbp]
	\centering
	\includegraphics{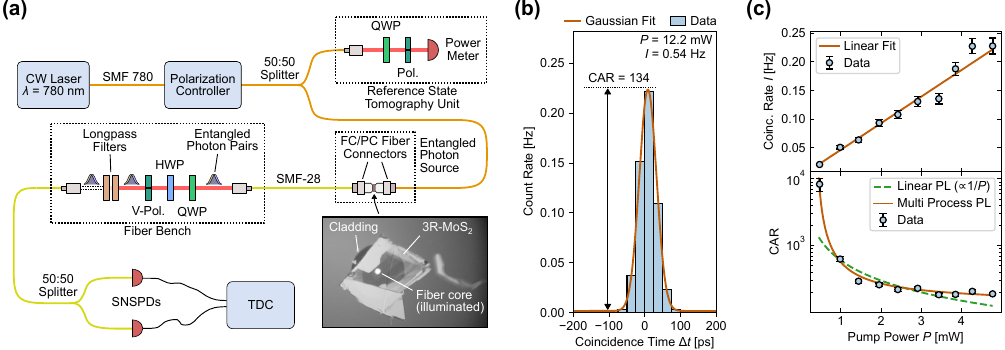}
	\caption{
		\textbf{Experimental Setup and SPDC Properties.}\\
		\textbf{(a)} Setup for the fully in-line generation and analysis of entangled photon pairs. The light of a 780 nm continuous wave (CW) laser is guided through a polarization controller and subsequently split towards a reference output and a 3R-MoS\textsubscript{2} crystal located between two fiber connectors. The generated photon pairs exit the SMF-28 single-mode fiber and are analyzed based on co-polarized state projections. The detection of entangled photon pairs is performed with superconducting nanowire single-photon detectors (SNSPDs), connected to an electronic time-to-digital converted (TDC).
		\textbf{(b)} Histogram of measured coincidence time differences for a high efficiency configuration of the setup.
		\textbf{(c)} Total net rate of coincidences $I$ for the peak around $\Delta t = 0$ and the coincidences-to-accidentals ratio (CAR) as a function of pump power $P$.
	}
	\label{fig:setup_and_powerdep}
\end{figure*}

We use the custom-built setup in Figure~\ref{fig:setup_and_powerdep}(a) for tuning and analyzing the generation of entangled photons in a fully in-line fiber geometry. In the setup, the light of a fiber-coupled continuous-wave (CW) laser at a wavelength of $\lambda = 780\,\mathrm{nm}$ is guided through a fiber-based polarization controller towards a 50:50 splitter. One half of the light is guided towards a reference output, where the reference polarization state $\ket{\psi_\mathrm{R}}$ is determined through a quarter-wave-plate- and linear-polarizer-based polarimetry scheme. The other half is guided towards a 3R-MoS\textsubscript{2} flake located on the end-facet of an FC/PC fiber connector, which we pre-selected based on second-harmonic generation efficiency for a fundamental wavelength of 1560 nm with a fs-laser. A detailed description of the crystal selection and sample fabrication procedure is presented in Section \ref{sec:meth:crys_select_and_sample_fab}.

A reflection microscope image of the crystal, recorded prior to the mating of both connectors, is shown in the bottom right of Figure~\ref{fig:setup_and_powerdep}(a). The image was recorded through a 780 nm bandpass filter and shows high transparency for the material over the illuminated core of the fiber. This indicates efficient coupling of the pump light into the high-index material and field enhancement due to Fabry-Perot interference effects, giving a boost to the SPDC conversion efficiency.

To measure the film thickness of the active region over the fiber core as accurately as possible, we performed a spectroscopic ellipsometry measurement of another flake obtained from the same bulk crystal, from which we derived the material's refractive index (Section \ref{sec:ellipso}). We then measured the transmission spectrum of the fiber network with the 3R-MoS\textsubscript{2} flake located between the two connectors after mating. Through modeling of the thickness-dependent material transmission in an optical transfer matrix scheme, we determined a material thickness of 408 nm for the sample (Figure~\ref{fig:crystal_thickness_fit}).

The photon pairs generated inside the flake couple into the connected SMF-28 fiber and propagate towards the output collimator. The light exiting the fiber is analyzed with co-polarized state projections by appropriately adjusting the rotation angles of the half-wave-plate (HWP) and quarter-wave-plate (QWP). After filtering out the remaining pump light, the photon pairs are coupled into another SMF-28 fiber for splitting and coincidence detection with super-conducting nanowire single-photon detectors (SNSPDs) connected to a time-to-digital converter (TDC). This post-processing hardware is fully compatible with fiber bench integration.

\subsection{SPDC Time Statistics and Power Dependence}
Tuning the detected photon state and the polarization controller to a high efficiency configuration, we obtain the histogram of coincidence times $\Delta t$ in Figure~\ref{fig:setup_and_powerdep}(b). The histogram shows a strong peak of coincidence times around $\Delta t = 0$, amounting to a total coincidence rate of $I = (0.539 \pm 0.009)\,\mathrm{Hz}$ above the background.

Fitting a Gaussian curve, we obtain a full-width-at-half-max (FWHM) of 57 ps, close to the $\sim$50 ps time resolution of our detection system. We attribute this ultra-short correlation time to two effects:
First, as phase-matching requirements are severely reduced, SPDC in 3R-MoS\textsubscript{2} thin films produces spectrally broadband two-photon states \cite{weissflogTunableTransitionMetal2024}. These correspond to very narrow distributions in the corresponding Fourier (i.e. time) domain. Second, in our experimental setup, the propagation distance of this broadband pulse inside optical fibers before detection is only on the order of 10 m, which limits temporal broadening due to dispersion.

Furthermore, we obtain a peak-to-background ratio of $134.1 \pm 2.0$ from the fit. As we lower the pump power (Figure \ref{fig:setup_and_powerdep}(c) bottom), the coincidences-to-accidentals ratio (CAR) increases substantially, up to a value of $(8.3\pm1.8)\times10^{3}$ at $P = 0.47\,\mathrm{mW}$.
This increase deviates substantially from the typical $\propto1/P$ behavior. Since we also observe the linear relationship between the coincidence rate and pump power that is characteristic for the SPDC process (Figure \ref{fig:setup_and_powerdep}(c) top), this deviation must arise from a nonlinear dependence of the background signal on $P$. This conclusion is further supported by an examination of the single count rates for each channel of the TDC (Figure~\ref{fig:single_countrates}), which show significant deviations from linearity at low power. As our pump wavelength of $\lambda = 780\,\mathrm{nm}$ corresponds to a photon energy of $E = 1.6\,\mathrm{eV}$ that lies below the 1.85 eV optical bandgap of 3R-MoS\textsubscript{2} \cite{xuCompactPhasematchedWaveguided2022}, a potential explanation for the nonlinearity could be strong contributions to the total photoluminescence (PL) from defect states, which saturate at high intensities.
This highlights the importance of reducing the defect concentration through improved crystal growth or annealing procedures.

To explicitly account for the nonlinear PL power dependence, we fit a multi-process PL model to the single count rates of the TDC channels (Figure \ref{fig:single_countrates}), which we use in our theoretical calculation of the CAR. Through this, we observe excellent agreement with the measured values.

Our obtained CAR value at low power exceeds the previously reported ones for SPDC in van der Waals materials
\cite{
	guoUltrathinQuantumLight2023,
	weissflogTunableTransitionMetal2024,
	fengPolarizationentangledPhotonpairSource2024,
	guoPolarizationEntanglementEnabled2024,
	trovatelloQuasiphasematchedDownconversionPeriodically2025,
	liangTunablePolarizationEntangled2025,
	lyuTunableEntangledPhotonpair2025,
	luCounterpropagatingEntangledPhoton2025,
	linNonlinearPhasematchedVan2026,
	joshiInLineFiberIntegratedPhotonPair2026a}
despite substantially greater pair generation efficiencies being observed in References \cite{
	guoUltrathinQuantumLight2023,
	fengPolarizationentangledPhotonpairSource2024,
	guoPolarizationEntanglementEnabled2024,
	trovatelloQuasiphasematchedDownconversionPeriodically2025,
	liangTunablePolarizationEntangled2025,
	lyuTunableEntangledPhotonpair2025,
	luCounterpropagatingEntangledPhoton2025,
	linNonlinearPhasematchedVan2026,
	joshiInLineFiberIntegratedPhotonPair2026a}.
We attribute this, in large part, to the nonlinear reduction in the background signal at low pump powers for our 3R-MoS\textsubscript{2} based source and, to a lesser degree, the excellent time resolution of our detection system and narrow width of the coincidence peak. Both effects reduce the background of accidental counts.

Furthermore, we note that the conversion efficiency of our photon pair source could be improved upon by selecting a 3R-MoS\textsubscript{2} crystal with greater thickness, to exploit the full 625 nm coherence length for down-conversion from 780 nm to 1560 nm.
Beyond the coherence length, substantial improvements can be achieved by stacking crystals in a periodically poled or twisted fashion, as it has been demonstrated in Reference \cite{trovatelloQuasiphasematchedDownconversionPeriodically2025} for 3R-MoS\textsubscript{2} and References \cite{linNonlinearPhasematchedVan2026, joshiInLineFiberIntegratedPhotonPair2026a} for rBN, respectively.

These findings highlight the importance of time resolution, dispersion management, and the role of the background signal for photon pair sources. The high signal-to-noise ratio enables us to implement the in-line, polarization-tunable state generation method, which we present in the following.

\subsection{Polarization State Mapping}

\begin{figure*}[htb]
	\centering
	\includegraphics{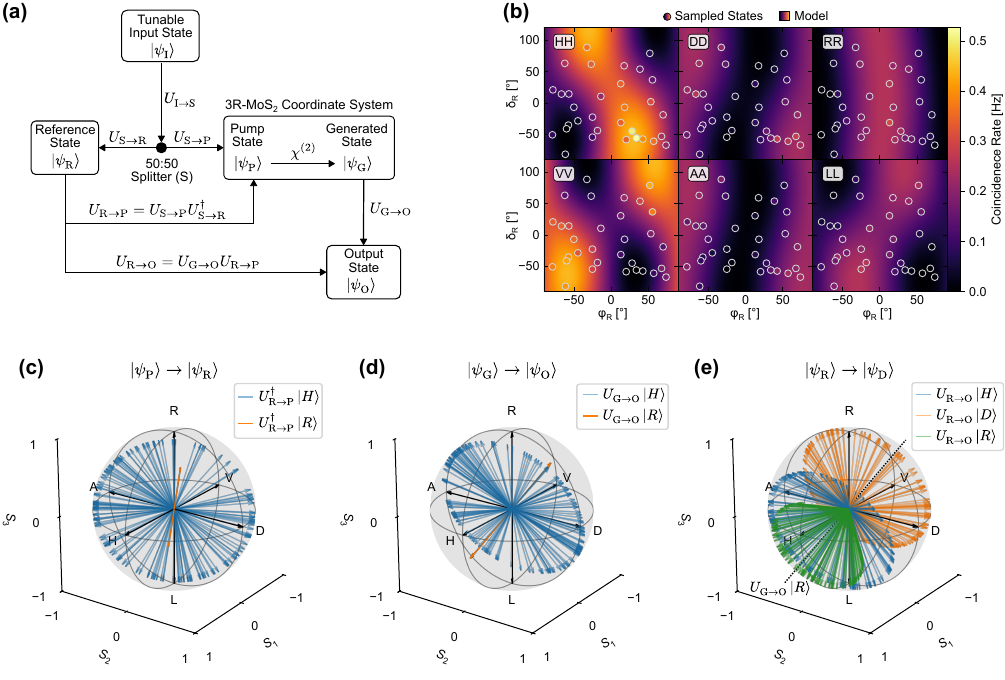}
	\caption{
		\textbf{Device Calibration and Polarization State Mapping.}
		\textbf{(a)} System of unitary transformations formed by the entangled photon generation setup.
		\textbf{(b)} Measured coincidence rates for each state projection of the sampled calibration states as a function of the corresponding reference state polarization angle $\varphi_\mathrm{R}$ and retardance $\delta_\mathrm{R}$.
		Based on these samples, $U_{\mathrm{R}\to\mathrm{P}}$ and $U_{\mathrm{G}\to\mathrm{O}}$ are obtained via least-squares optimization. $U_{\mathrm{R}\to\mathrm{P}}$ and $U_{\mathrm{G}\to\mathrm{O}}$ are then used for calculating the output state. The color maps show the expected coincidence rates of the modeled output state as a function of $\varphi_\mathrm{R}$ and $\delta_\mathrm{R}$.
		\textbf{(c)-(e)} State mappings highlighting the partial uniqueness of the obtained optima for $U_{\mathrm{R}\to\mathrm{P}}$ and $U_{\mathrm{G}\to\mathrm{O}}$. All optimized $U_{\mathrm{R}\to\mathrm{P}}$ map the same reference state onto either $\ket{R}$ or
		$\ket{L}$, while the semi-orthogonal states vary freely. Similarly, all $U_{\mathrm{G}\to\mathrm{O}}$ always map $\ket{R}$ or $\ket{L}$ onto the same state. For their product, all optimal solutions map the same states onto cones around the axis of $U_{\mathrm{G}\to\mathrm{O}} \ket{R}$. 
	}
	\label{fig:spdc_calib}
\end{figure*}

To use in-plane $\chi^{(2)}$ tensor components of 3R-MoS\textsubscript{2} for tunable state generation, polarization control of the pump radiation and generated photon pairs is critical.
However, as briefly addressed above, the unavoidable bends and twists of optical fibers in practical deployment create anisotropic stress patterns that induce optical birefringence in the material. Thus, in practice, any sufficiently long fiber represents an arbitrary sequence of birefringent materials \cite{hechtUnderstandingFiberOptics2015}.
For each wavelength of light propagating through the fiber, this sequence represents an elliptical retarder, which, depending on its specific properties, can transform a single polarization state onto any other. Crucially, this transformation is unitary, and preserves the scalar product of the mapped polarization states. In the Jones formalism, unitary transformations are represented by unitary matrices and can be generated, for example, with the parametrization:
\begin{align}
    U = \begin{pmatrix}
        \overline{\alpha} & -\beta \\
        \overline{\beta} & \alpha \\
    \end{pmatrix}, \quad |\alpha|^2 + |\beta|^2 = 1,
\end{align}
where $\alpha$ and $\beta$ are complex numbers and $\overline{\alpha}$,  $\overline{\beta}$ are their complex conjugates. Note that, in general, the complex transpose $U^\dagger$ of $U$ is its inverse (and also unitary). Due to the normalization constraint, $U$ is fully characterized by just three real-valued parameters. This makes the practical determination of $U$ feasible with only a limited number of test polarization states injected into the fiber. 

Figure~\ref{fig:spdc_calib}(a) depicts a schematic representation of the unitary transformations between the polarization states at key points in the fiber network of our experimental setup.
Note that we verified the compatibility of the 50:50 fiber splitter with a unitary model, i.e. polarization insensitivity, with a series of test measurements between the reference output and the end facet of the SMF 780 fiber, prior to the connection to the SMF-28 fiber (Figure~\ref{fig:ref_to_sample_state_mapping}).

The reference and pump states are each determined by a different chain of unitary transformations acting upon the tunable input state. As any chain of unitary transformations is itself a unitary transformation, both polarization states are also linked to each other by a unitary transformation, which we denote $U_{\mathrm{R}\to\mathrm{P}}$. 

For the generated pair, the unitary transformation $U_{\mathrm{G}\to\mathrm{O}}$ of the SMF-28 fiber acts independently on the signal and idler photons. The corresponding transformation acting on the two-photon polarization state is therefore $U_{\mathrm{G}\to\mathrm{O}}^\mathrm{s} \otimes U_{\mathrm{G}\to\mathrm{O}}^\mathrm{i}$.
In principle, both $U_{\mathrm{G}\to\mathrm{O}}^\mathrm{s}$ and $U_{\mathrm{G}\to\mathrm{O}}^\mathrm{i}$ can depend on the specific signal and idler wavelength components. However, in practice, we observed no substantial deviations under the assumption of a single, wavelength-independent $U_{\mathrm{G}\to\mathrm{O}}$ for both channels.

To obtain a suitable pair of $U_{\mathrm{R}\to\mathrm{P}}$ and $U_{\mathrm{G}\to\mathrm{O}}$, we use the polarization controller to generate a series of 32 random input polarization states. For each of these, we record the coincidence rates for the co-polarized state projections of $\ket{HH}$, $\ket{DD}$, $\ket{RR}$, and their orthogonal counterparts, as well as the reference polarization state. Note that we define the diagonal polarization state as $\ket{D} = \frac{1}{\sqrt{2}}(\ket{H} + \ket{V})$.

Figure \ref{fig:spdc_calib}(b) shows the measured coincidence rates for each of the generated states in a color scaled scatter plot. The values are displayed as a function of the retardance $\delta_\mathrm{R}$ and polarization angle $\varphi_\mathrm{R}$ of the recorded reference polarization state.

To derive $U_{\mathrm{R}\to\mathrm{P}}$ and $U_{\mathrm{G}\to\mathrm{O}}$,
we model the polarization state evolution and SPDC process as a function of $\ket{\psi_\mathrm{R}}$ in the Jones formalism:
\begin{align}
	\ket{\psi_\mathrm{O}} = I_0 (U_{\mathrm{G}\to\mathrm{O}} \otimes U_{\mathrm{G}\to\mathrm{O}}) \chi^{(2)} U_{\mathrm{R}\to\mathrm{P}} \ket{\psi_\mathrm{R}}, \label{eq:fiber_spdc}
\end{align}
where $I_0$ is the maximum detectable coincidence rate and $\chi^{(2)}$ is chosen to be real-valued and norm-preserving:
\begin{align}
	\chi^{(2)} = \frac{1}{\sqrt{2}} \begin{pmatrix}
		1 & 0  \\
		0 & 1  \\
		0 & 1  \\
		-1 & 0 \\
	\end{pmatrix}.
\end{align}
Based on Equation~\eqref{eq:fiber_spdc}, we derive $I_0$ and the unitary matrix components in a least-squares optimization scheme, minimizing the total coincidence rate error for all sampled states and state projections.
The model coincidence rates for each analyzer state are displayed in Figure~\ref{fig:spdc_calib}(b) alongside the sampled values.
Furthermore, by attempting the optimization many times with slightly varied initial parameters, we obtain many different pairs of unitary transformations which fit the data optimally. In Figure~\ref{fig:spdc_calib}(c)-(e), we have displayed the Stokes parameters of several key state mappings for their examination.

We find that $U_{\mathrm{R}\to\mathrm{P}}^\dagger$, the inverse of the first unitary, consistently maps the state of right circular polarization $\ket{R}$ in the coordinate system of the 3R-MoS\textsubscript{2} crystal to one of two orthogonal states, near $\ket{R}$ or $\ket{L}$ in the reference coordinate system.
However, the image of the linear polarization states, as shown for $U_{\mathrm{R}\to\mathrm{P}}^\dagger \ket{H}$, varies freely on a circle perpendicular to those states. The same general behavior applies to the second unitary transformation $U_{\mathrm{G}\to\mathrm{O}}$ with a different pair of image states for $\ket{R}$/$\ket{L}$. Additionally, we find that $U_{\mathrm{R}\to\mathrm{O}}$, the product of $U_{\mathrm{R}\to\mathrm{P}}$ and $U_{\mathrm{G}\to\mathrm{O}}$, consistently maps the same reference states onto cones around $U_{\mathrm{G}\to\mathrm{O}}\ket{R}$, which is a consequence of the scalar product preserving nature of unitary transformations.

We emphasize here that the non-uniqueness of the unitary transformations with respect to the linear polarization states is not merely a consequence of the co-polarized analyzer scheme that could be resolved in a full two-photon polarization state tomography. If the true transformation $U_{\mathrm{G}\to\mathrm{O}}$ and crystal orientation are known, the generated state can be fully characterized in a co-polarized tomography scheme based on just three measurements (Section~\ref{sec:copol_tomo}). Instead, the non-uniqueness is a consequence of the $\chi^{(2)}$ tensor symmetries of 3R-MoS\textsubscript{2} and the freedom to choose any orthogonal pair of the linear polarization states in the crystal coordinate system as logical states $\ket{0}$ and $\ket{1}$ (Section~\ref{sec:basis_choice}). Any combination of optimal transformations is therefore a valid choice for the tunable generation of entangled photon pairs.

\subsection{Polarization State Tuning}
Choosing one of the pairs of optimal unitary matrices, we define $U_{\mathrm{G}\to\mathrm{O}} \ket{H}$ and $U_{\mathrm{G}\to\mathrm{O}} \ket{V}$ as logical $\ket{0}$ and $\ket{1}$ states, respectively. Then, the form of the produced two-photon output state $\ket{\psi_\mathrm{O}}$ is analogous to that of the generated state in Equation~\eqref{eq:gen_biph_state} in the $\ket{0}$ and $\ket{1}$ basis.

To obtain any arbitrary combination of $\varphi_\mathrm{P}$ and $\delta_\mathrm{P}$, i.e. pump state $\ket{\psi_\mathrm{P}}$, we calculate the corresponding reference state
$\ket{\psi_\mathrm{R}} = U_{\mathrm{R}\to \mathrm{P}}^\dagger \ket{\psi_\mathrm{P}}$
and rotate the QWP and polarizer of the reference output (Figure \ref{fig:setup_and_powerdep}(a), top-right) to detect this state with minimal power. We then adjust the paddle positions of the polarization controller to minimize the measured reference power in a fully-autonomous Bayesian Optimization scheme \cite{nogueriaBayesOptPython2014}. Through this method, we achieve typical extinction ratios of $P_\mathrm{max} / P_\mathrm{min} = 97$, where $P_\mathrm{max}$ and $P_\mathrm{min}$ are the maximum-detectable and minimized-to power values, respectively, and typical obtained-to-target-state scalar products of 98\%. 

\begin{figure*}[htbp]
	\centering
	\includegraphics{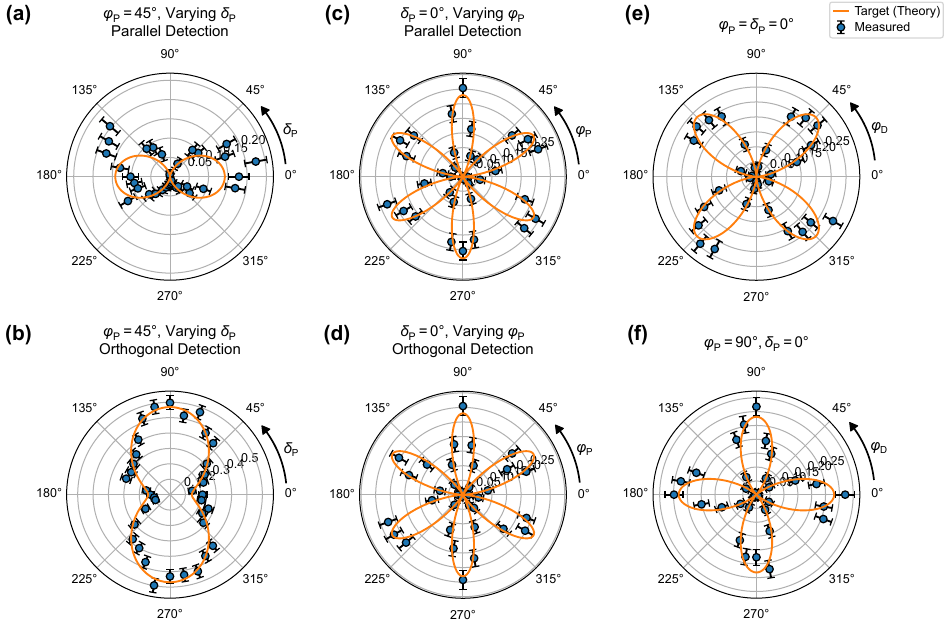}
	\caption{\textbf{Tuning of the Generated Two-Photon Polarization State.}
		\textbf{(a)}, \textbf{(b)} Measured and theoretically expected coincidence rates when varying the retardance $\delta_\mathrm{P}$ from 0° to 360° while $\varphi_\mathrm{P} = 45\degree$, for parallel and orthogonal detection, respectively. The theoretical curves have been scaled to both measured datasets in a joint least-squares optimization with a single parameter.
		\textbf{(c)}, \textbf{(d)} Likewise, coincidence rates for the case of varying the polarization angle $\varphi_\mathrm{P}$ from 0° to 360° while $\delta_\mathrm{P} = 0\degree$.
		\textbf{(e)}, \textbf{(f)} Coincidence rates for fixed pump states $\ket{H}$ ($\varphi_\mathrm{P} = 0\degree$) and $\ket{V}$ ($\varphi_\mathrm{P} = 90\degree$), respectively, varying the polarization angle $\varphi_\mathrm{D}$ of the detected state, while $\delta_\mathrm{D} = 0$.
	}
	\label{fig:state_sweeps}
\end{figure*}

To verify the validity and practicality of deterministic state generation, we perform a variety of polarization-dependent coincidence rate measurements.
In Figure~\ref{fig:state_sweeps}(a) and (b), we simultaneously vary the retardance of $\ket{\psi_\mathrm{P}}$ and the detected analyzer state $\ket{\psi_\mathrm{D}}$, which is defined analogously to the pump polarization state in Equation~\eqref{eq:pump_state}. The resulting coincidence rates are in good agreement with the calculated theoretical values that have been fitted to the data with a single scaling parameter. Most notably, for the case of $\delta_\mathrm{P} = \pm90\degree$, i.e. $\ket{\psi_\mathrm{P}} = \ket{\tilde{R}}$ and $\ket{\psi_\mathrm{P}} = \ket{\tilde{L}}$, the fully co-polarized states
$\ket{\tilde{L}\tilde{L}}$ and $\ket{\tilde{R}\tilde{R}}$ are produced, respectively. Note that $\ket{\tilde{L}}$ and $\ket{\tilde{R}}$ are defined analogously to their $\ket{H}$ and $\ket{V}$ basis counterparts.
Correspondingly, maximum coincidence rates are measured at these points in the case of orthogonal detection, and full extinction is achieved for parallel detection.

Additionally, we set the pump retardance to zero and vary the polarization angles $\varphi_\mathrm{P}$ and $\varphi_\mathrm{D}$ simultaneously to probe state generation along the whole manifold of maximum entanglement.
We obtain the same 6-fold-symmetric pattern for both parallel and perpendicular detection. Again, both results are in good agreement with the theoretical values. Note that, in contrast to second-harmonic generation measurements, the 6-fold patterns are not shifted by 30° with respect to each other~\cite{weissflogTunableTransitionMetal2024}.

\begin{figure*}[htbp]
	\centering
	\includegraphics{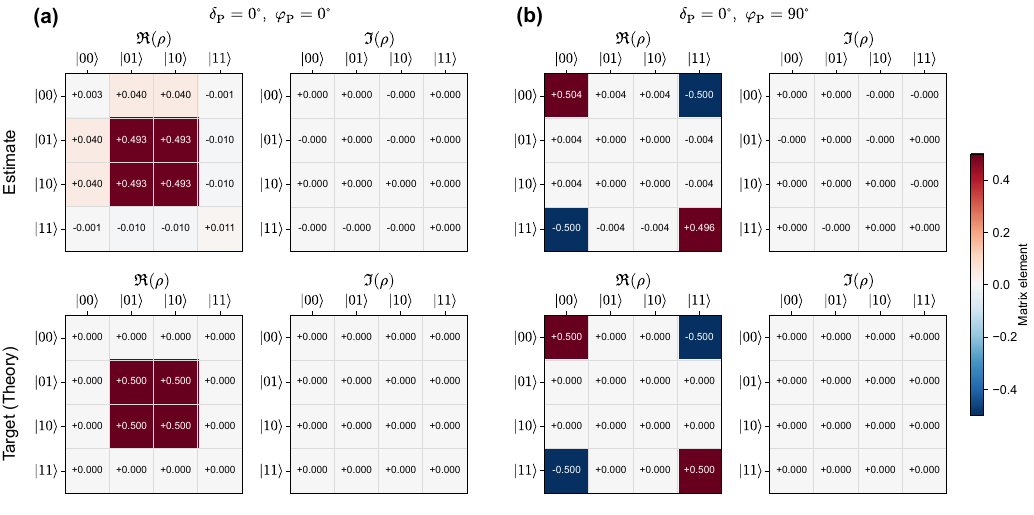}
	\caption{
		\textbf{Density Matrix Estimation.}
		\textbf{(a)} Estimated density matrix $\rho$ of the output state, obtained via least-squares optimization based on the co-polarized projection measurement results for horizontal pump ($\delta_\mathrm{P} = 0$, $\varphi_\mathrm{P} = 0$) and density matrix of the target Bell state $\ket{\psi^+}$.
		\textbf{(b)} Likewise, $\rho$ for vertical pump polarization ($\delta_\mathrm{P} = 0$, $\varphi_\mathrm{P} = 90\degree$) and target Bell state $\ket{\phi^-}$.
	} 
	\label{fig:est_dens_mat}
\end{figure*}

Finally, to explicitly verify the generation of the Bell states $\ket{\psi^+}$ and $\ket{\phi^-}$, we set $\varphi_\mathrm{P}$ to 0° and 90°, respectively, and vary $\varphi_\mathrm{D}$. The respective measured coincidence rates show clear 4-fold symmetry as a function of $\varphi_\mathrm{D}$ and are offset to each other by 45°. As previous free-space SPDC measurements of rBN have shown \cite{liangTunablePolarizationEntangled2025}, these are the exact signatures one expects for $\ket{\psi^+}$ and $\ket{\phi^-}$. To examine the quality of the output states, we perform a maximum likelihood estimation of the density matrix $\rho$ \cite{banaszekMaximumlikelihoodEstimationDensity1999} based on the co-polarized projection measurement results in Figure~\ref{fig:state_sweeps}(e) and (f). Using each of the corresponding Bell states as an initial guess, we obtain the real and imaginary parts of $\rho$ displayed in Figure~\ref{fig:est_dens_mat}. Note that, here, we rely on prior knowledge of the approximate two-photon polarization state, as co-polarized projection measurements are not sufficient to uniquely resolve the general case. For both states, the estimated density matrix is in decent agreement with the theoretical target.

Additionally, we estimate the state purity $P$, concurrence $C$, and fidelity to the target state $F$ in a Monte Carlo scheme, based on the uncertainties of the coincidence rate data points. Through the evaluation of $N = 1000$ samples, we obtain an average purity $P = 97.8_{-1.7}^{+1.9}\,\%$, concurrence $C = 97.11_{-1.5}^{+2.2}\,\%$, and fidelity $F = 98.3_{-0.8}^{+1.1}\,\%$ for the state generated with horizontal pump polarization $\varphi_\mathrm{P} = 0$. Note that, for each quantity, the superscript and subscript $+$ and $-$ correspond to the 84th and 16th percentiles of the distribution (analogous to one symmetric standard deviation). Likewise, we obtain $P = 99.7_{-0.3}^{+0.3}\,\%$, $C = 99.3_{-0.4}^{+0.7}\,\%$, and $F = 99.3_{-0.2}^{+0.3}\,\%$ for the case of vertical pump polarization $\varphi_\mathrm{P} = 90\degree$.
These results indicate that, especially for the latter case, the output two-photon state is of decent quality, despite the broadband nature of the generated photon pairs and potential chromatic variations in $U_{\mathrm{G}\to\mathrm{O}}$. Further improvements in these metrics can, most likely, be achieved when periodically-poled or twisted stacks of vdW crystals
\cite{
	trovatelloQuasiphasematchedDownconversionPeriodically2025,
	linNonlinearPhasematchedVan2026,
	joshiInLineFiberIntegratedPhotonPair2026a},
with higher SPDC efficiency are used as nonlinear media, as the spectral bandwidth naturally decreases with rising interaction length, due to phase-matching constraints.

Overall, these results strongly indicate that, even in our greatly simplified, co-polarized tomography scheme and without knowledge of the crystal orientation, one is able to deterministically generate the wide selection of two-photon polarization states intrinsically compatible with the in-plane $\chi^{(2)}$ tensor elements of 3R-MoS\textsubscript{2} and similar materials.

\section{Conclusion}
We have demonstrated a fully in-line, fiber-based source of polarization-tunable entangled photon pairs based on SPDC in a sub-micron 3R-MoS\textsubscript{2} thin film transferred onto a standard fiber connector and a method to control its state in the presence of fiber birefringence. By combining the intrinsic $C_{3v}$ symmetry of the in-plane $\chi^{(2)}$ tensor with a calibrated unitary description of the fiber network, the polarization state generated at the nonlinear material can be deterministically addressed through a reference polarization measurement at the pump wavelength. This enables practical control of the two-photon output state without free-space manipulation, crystal-orientation readout, or access to the material after connector mating. For future practical applications, the laser source, polarization control and reference measurement can be combined in a single device.

The source generates photon pairs in the near-infrared telecommunication bands with a narrow coincidence peak limited by the timing resolution of our detection system and a coincidence-to-accidentals ratio reaching $(8.3\pm1.8)\times10^{3}$ due to a nonlinear drop of the photoluminescence background at low pump powers. Polarization-resolved coincidence measurements confirm the predicted dependence of the generated state on the pump polarization, including the generation of co-polarized photon pairs for circular pump states and the Bell states $\ket{\psi^+}$ and $\ket{\phi^-}$ for orthogonal choices of linear pump polarization.

These results establish fully fiber-integrated 3R-MoS\textsubscript{2} as a compact and tunable platform for telecom-band entangled photon-pair generation. Further improvements in brightness and noise performance should be achievable through optimized crystal thickness, quasi-phase-matched or twisted stacks, and, potentially, reduced defect-related photoluminescence. More broadly, the approach shows how the symmetry of van der Waals nonlinear materials can be combined with calibrated fiber polarization control to realize robust, alignment-free quantum light sources compatible with existing optical-fiber infrastructure.

\section{Methods} \label{sec:meth}
\subsection{Crystal Selection and Sample Fabrication} \label{sec:meth:crys_select_and_sample_fab}
A 3R-MoS\textsubscript{2} bulk crystal was obtained from HQ Graphene. Flakes were exfoliated from the bulk material using adhesive tape and transferred onto polydimethylsiloxane (PDMS) films (Gel-Pak, PF-30-X4) by pressing the tape onto the material and peeling it off. Candidate flakes were selected based on their optical uniformity and lateral size and evaluated with respect to their second-harmonic generation efficiency produced in transmission under excitation with a fs-pulsed-laser (FWHM $\sim$200 fs) at a wavelength of 1560 nm (Light Conversion Carbide (Pump) and Orpheus (Optical Parametric Amplifier) combination).

After pre-selection, the chosen flake was transferred onto the core of an SMF-28 fiber patch cable (Thorlabs P1-SMF28E-FC-1), exposed through the facet of an FC/PC fiber connector in a dry-transfer stamping procedure. After transfer, a reflection image of the connector with the attached flake was recorded with a custom-built microscope using a halogen lamp (Thorlabs OSL2IR) and 780 nm bandpass filter (Thorlabs FBH780-10). A SMF-780 patch cable with FC/PC connectors (Thorlabs P1-780A-FC) was then mated with the SMF-28 fiber in a high-precision mating sleeve (Thorlabs ADAFCPM2), placing the 3R-MoS\textsubscript{2} flake between the two fiber cores.

\subsection{Fiber-Based SPDC Setup}
The pump light was provided by a fiber-coupled, grating-stabilized CW laser at $\lambda=780\,\mathrm{nm}$ (Toptica DL Pro). The pump polarization was controlled with a motorized fiber polarization controller (Thorlabs MPC320) before being sent to a 50:50 fiber splitter (Thorlabs PN780R2F1). One output of the splitter was used as a reference arm for pump-polarization readout using a quarter-wave plate (QWP), linear polarizer, and optical power meter (Thorlabs AQWP10M-980, LPNIRE100-B, S130C, respectively) while the other output was connected to the patch cable with the 3R-MoS\textsubscript{2} sample.
Specifically for the power-dependent coincidence measurements, we added a variable attenuator consisting of a half-wave plate (HWP) and linear polarizer in a fiber-bench platform (Thorlabs FBR-AH2, FBR-LPNIR, FB-76W, respectively) after the laser.

Photon pairs generated in the flake were coupled into the connected SMF-28 fiber and guided to the analysis stage. The output light was collimated with an achromatic fiber-coupler (Thorlabs PAF2-A7C) and passed through a polarization analyzer consisting of an achromatic QWP, HWP, and linear polarizer (Thorlabs AQWP05M-1600, AHWP05M-1430, WP25M-UB1, respectively). Residual pump light and, partly, photoluminescence were suppressed with a combination of a 900 nm (Edmund Optics 66-229) and 1500 nm (Thorlabs FELH1500) longpass filters. The filtered photons were coupled into another SMF-28 fiber with the same model of fiber coupler used for collimation and split probabilistically with a broadband 50:50 fiber splitter (Thorlabs TW1550R5F1) before detection with superconducting nanowire single-photon detectors (SNSPDs) (Single Quantum Eos). The individual counts produced by the SNSPDs were recorded with a time-to-digital converter (TDC) (qutools quTAG HR) post-processed with a custom-written Python script to determine coincidence time differences.

\subsection{Ellipsometry}
The refractive index of 3R-MoS\textsubscript{2} was determined by variable-angle spectroscopic ellipsometry (VASE) on a large-scale reference flake exfoliated from the same bulk crystal as the device flake. The measurement was performed with a commercial Woolam M-2000D ellipsometer in Mueller matrix mode with its focusing probes attached over a wavelength range of 192 nm - 1688 nm and at incidence angles of 45° to 65° in 5° steps. The recorded ellipsometric parameters were fitted with a multi Tauc-Lorentz oscillator model using the manufacturer-provided CompleteEASE software to extract the wavelength-dependent complex refractive index for the in-plane and out-of-plane directions.

\subsection{Thickness Characterization}
To determine the thickness of the 3R-MoS\textsubscript{2} flake in the region located over the fiber core after connector mating, we coupled the light from the halogen lamp into another SMF-780 patch cable to create a broadband fiber-based source of illumination. The light source was then coupled into the SPDC setup, in place of the CW laser. To record the transmission spectrum, the free-space output of the SMF-28 patch cable was connected to a multi-mode fiber. The output of the multi-mode fiber was then collimated and guided towards a free-space CCD-sensor-based digital spectrometer (Horiba iHR320). A reference measurement of the light source with equal exposure and gain settings was obtained by connecting its output directly to the multi-mode fiber. All measurement results were corrected by subtracting the results of a dark measurement with the light-source turned off.

The transfer-matrix method previously described in Reference \cite{abtahiThicknessDependenceLinear2026} was used to model the wavelength- and thickness-dependent transmission of the 3R-MoS\textsubscript{2} film between the two fused silica fiber cores. The film thickness was obtained by least-squares fitting of the modeled transmission to the measured spectrum, where the slowly varying attenuation of the fiber network was included as a spline with three support points.

\section*{Acknowledgments}
The authors acknowledge the support from the Deutsche Forschungsgemeinschaft (DFG, German Research Foundation), Project ID: 437527638—IRTG 2675 (Meta-Active), Collaborative Research Center (CRC/SFB) 1375 NOA Project B3, and the ARC Centres of Excellence program (CE200100010).
This work used the ACT node of the NCRIS-enabled Australian National Fabrication Facility (ANFF).

\section*{Disclosures}
The authors declare no conflicts of interest.

\section*{Data Availability}
Data underlying the results presented in this paper are not publicly available at this time but may be obtained from the authors upon reasonable request.

\section*{Author Contributions}
B.L. developed the concept, fabricated the samples, constructed and developed software for the setup, performed the coincidence measurements, performed the data analysis and wrote the manuscript. F.A. assisted in the construction of the setup, sample fabrication and coincidence measurements. D.C. performed the ellipsometry measurement. D.N. and F.E. supervised B.L. and F.A., provided funding and resources for the project, and suggested revisions to the manuscript. All authors proofread the manuscript.

\printbibliography

\end{multicols}

\newpage
\appendix

\renewcommand{\thesection}{S\arabic{section}}
\renewcommand{\thesubsection}{S\arabic{section}.\arabic{subsection}}
\renewcommand{\thefigure}{S\arabic{figure}}
\renewcommand{\thetable}{S\arabic{table}}
\renewcommand{\theequation}{S\arabic{equation}}
\setcounter{section}{0}
\setcounter{figure}{0}
\setcounter{table}{0}
\setcounter{equation}{0}

\changefontsize{11pt}
\begin{adjustwidth}{0.6cm}{0.6cm}
	\captionsetup{width=\linewidth}
	\begin{center}
		\noindent{\textbf{\huge{Supplementary}}}\\[0.5cm]
	\end{center}
	\section{Ellipsometry of 3R-MoS\texorpdfstring{\textsubscript{2}}{2}} \label{sec:ellipso}
As described in the methods section of the main text, we have performed a variable-angle spectroscopic ellipsometry (VASE) measurement of a reference 3R-MoS\textsubscript{2} flake on a flat, silicon dioxide substrate that was exfoliated from the same bulk crystal as our sample for SPDC. Figure \ref{fig:ellipso_height_and_refr_index}(a) displays a height map of this flake, which we obtained via vertical scanning interferometry (VSI) with a white light interferometer (Bruker Contour GT). Notably, the sample has a large region of interest, suitable for ellipsometry, with a narrow, nearly normal thickness distribution with a mean of 1.87 µm and standard deviation of 17 nm.

\begin{figure}[htbp]
	\centering
	\includegraphics{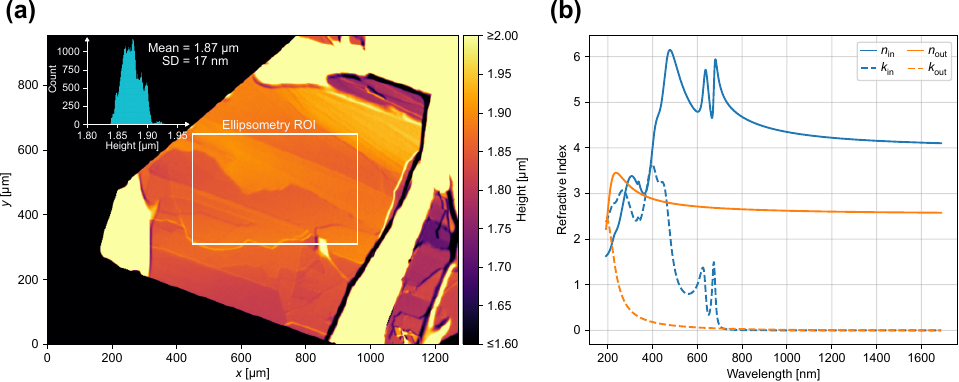}
	\caption{
		\textbf{(a)} Height map of the reference 3R-MoS\textsubscript{2} flake for the ellipsometry measurement obtained via vertical scanning interferometry (VSI). The flake features a large region of interest (ROI) of fairly homogeneous thickness, which was used for the ellipsometry measurement. The inset shows the distribution of thickness values inside the ROI with a mean value of $1.87\,\upmu\mathrm{m}$ and a standard deviation (SD) of 17 nm.
		\textbf{(b)} In-plane $n_\mathrm{in}$ and out-of-plane $n_\mathrm{out}$ refractive indices of 3R-MoS\textsubscript{2} obtained from the ellipsometry measurement. The optimized parameters of the Tauc-Lorentz based dispersion model are listed in Table~\ref{tab:tauc_lorentz_params}.
	}
	\label{fig:ellipso_height_and_refr_index}
\end{figure}

By performing the VASE measurement in Mueller matrix mode, and by fitting a multi Lorentz-Tauc oscillator dispersion model in addition to correction parameters for the sample non-uniformity, surface roughness, and potential angle of incidence offset to the data with the manufacturer-provided CompleteEase software, we obtain the in-plane $n_\mathrm{in}$ and out-of-plane $n_\mathrm{out}$ refractive index values of the material, which are displayed in Figure~\ref{fig:ellipso_height_and_refr_index}(b). The optimized oscillator parameters, where the contribution $\Delta\varepsilon_2$ of each oscillator to the total imaginary part of the dielectric function $\varepsilon_2$ is given by \cite{jellisonParameterizationOpticalFunctions1996}
\begin{align}
	\Delta\varepsilon_2 = 
	\begin{dcases}
		\frac{1}{E} \frac{A E_0 C (E - E_\mathrm{g})^2}{(E^2 - E_0^2)^2 + C^2 E^2}  & E > E_\mathrm{g} \\
		0 & E \leq E_\mathrm{g}
	\end{dcases}
\end{align}
are listed in Table~\ref{tab:tauc_lorentz_params}. Note that we did not achieve full numerical convergence for all individual parameters, and have, thus, omitted their uncertainties. Nonetheless, we observe three prominent peaks in $n_\mathrm{in}$ and $k_\mathrm{in}$, which are caused by excitonic resonances of the material \cite{xuCompactPhasematchedWaveguided2022}. Furthermore, compared to previous results \cite{xuCompactPhasematchedWaveguided2022}, we derive substantially greater values of $n$ and $k$ associated with the C-exciton, around the wavelength of $\lambda = 460\,\mathrm{nm}$.

\begin{table}[htbp]
  \centering
  \caption{Optimized oscillator parameters for the dispersion model of 3R-MoS\textsubscript{2}}
  \label{tab:tauc_lorentz_params}
  \begin{tabular}{lrrrr}
    \toprule
    Index & $A$ & $C$ & $E_0$ (eV) & $E_\mathrm{g}$ (eV) \\
    \midrule
    \multirow{9}{*}{$n_\mathrm{in}$, $\varepsilon_{\infty} = 0$}
      & 63.4851 & 0.044 & 1.837 & 1.648 \\
      & 697.3562 & 0.085 & 1.954 & 1.889 \\
      & 53.5901 & 0.397 & 2.699 & 1.524 \\
      & 35.8926 & 0.477 & 3.054 & 1.583 \\
      & 1.8749 & 0.088 & 3.689 & 2.858 \\
      & 49.2970 & 1.627 & 4.353 & 1.419 \\
      & 15.2957 & 0.776 & 5.460 & 3.954 \\
      & 60.9484 & 8.945 & 8.256 & 1.331 \\
    \midrule
    $n_\mathrm{out}$, $\varepsilon_{\infty} = 2.431$
    & 37.4798 & 2.110 & 6.044 & 0.824\\
    \bottomrule
  \end{tabular}
\end{table}

In Figure~\ref{fig:mm_elem} we have displayed the measured and modeled values of the Mueller matrix elements. Due to the out-of-plane orientation of the ordinary axis, the crystal preserves the polarization of purely TE- or TM-polarized light and the Stokes parameter $S_1$ ($H$/$V$-pol.) is not coupled to $S_2$ ($D$/$A$-pol.) or $S_3$ ($R$/$L$-pol.). Thus, $M_{11} = 1$, while all of the corresponding cross-coupling elements are zero.

\begin{figure}[htbp]
	\centering
	\includegraphics[scale=0.6]{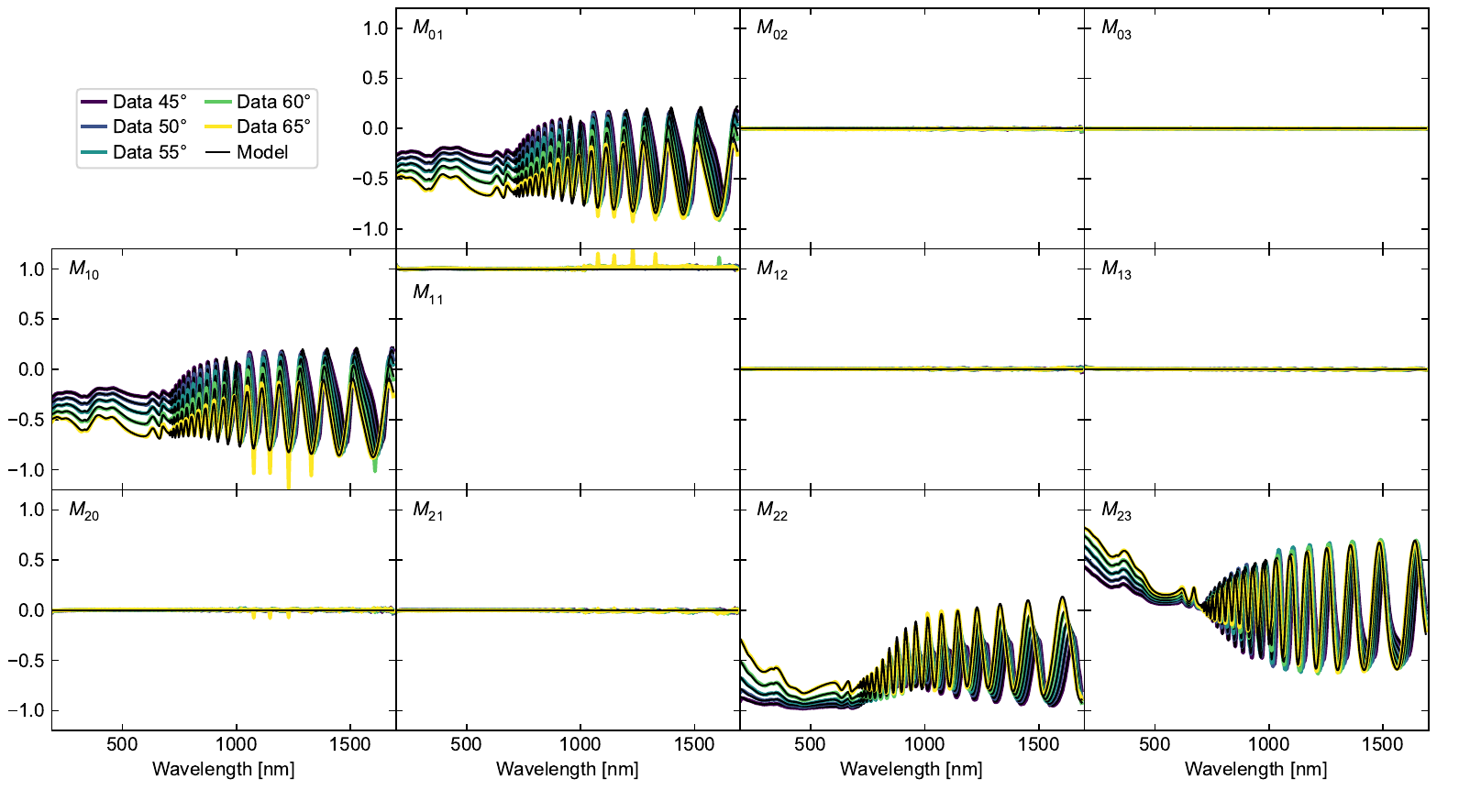}
	\caption{
		Measured and calculated Mueller matrix elements obtained from the ellipsometry measurement of the reference flake and fitted model, respectively. In addition to the oscillator parameters listed in Table~\ref{tab:tauc_lorentz_params}, the model fit yields optimized values for surface roughness $\rho = 1.69\,\mathrm{nm}$, angle of incidence offset $\delta = 0.591\degree$, and sample thickness non-uniformity $\tau = 3.46\%$.
	}
	\label{fig:mm_elem}
\end{figure}

For the remaining elements, we observe a large number of oscillations above $\lambda \approx 700\,\mathrm{nm}$. At this wavelength, the material becomes sufficiently transparent for thin-film interference effects to appear for the material thickness of $t \approx 1.87\,\upmu\mathrm{m}$.

\section{Spectroscopic Thickness Characterization}
Having obtained the refractive index of our 3R-MoS\textsubscript{2} material, we are able to determine the thickness of the material based on thin-film interference effects in the transmission spectrum. In Figure~\ref{fig:crystal_thickness_fit}(a), we have sketched the changes made to the fiber-based SPDC setup to allow probing of the specific region over the fiber core in situ.

\begin{figure}[htbp]
	\centering
	\includegraphics{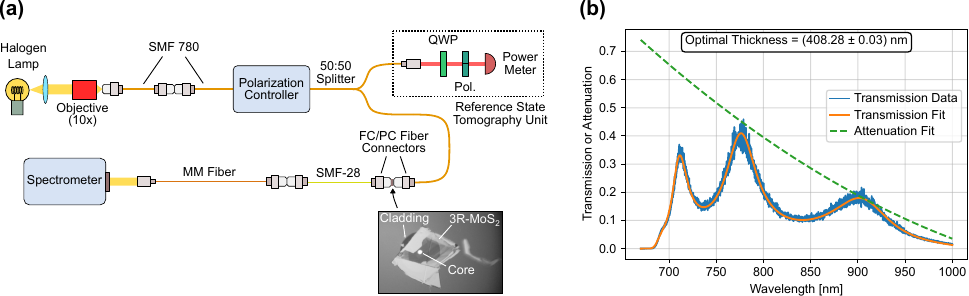}
	\caption{
		\textbf{Spectroscopic crystal thickness characterization.}\\
		\textbf{(a)} Modified fiber-based SPDC setup for measuring the transmission spectrum of the fiber network.
		\textbf{(b)} Recorded and fitted transmission spectra of the fiber network through the 3R-MoS\textsubscript{2} crystal branch. The general attenuation of the fiber network (mostly caused by the 50:50 splitter) was fitted as a spline with three support points. The optimal material thickness is determined as $(408.28 \pm 0.03)\,\mathrm{nm}$. 
	}
	\label{fig:crystal_thickness_fit}
\end{figure}

First, we create a broadband SMF-780-based light source by coupling the light of a halogen lamp into the fiber with a 10$\times$ magnification objective (Mitutoyo Plan Apo NIR). The overall coupling efficiency for this is quite low, but sufficient when detected with a sensitive spectrometer. This light source is then coupled into the fiber network of the setup in place of the laser.

The light leaving the sample branch through the SMF-28 fiber is then collected by connecting this fiber end with a multi-mode fiber. Note that, as the fiber core of the multi-mode fiber is substantially larger, the resulting coupling efficiency is consistent across repetitions. The light exiting the multi-mode fiber is then collimated and directed into a free-space-coupled digital spectrometer. To obtain a reference spectrum of the light source, the initial SMF-780 fiber is connected directly to the multi-mode fiber leading to the spectrometer.

We then model the transmission of the fiber network as the product
\begin{align}
	T(\lambda, t) = T_\mathrm{flake}(\lambda, t) A(\lambda, y_i)
\end{align}
where the transmission of the flake $T_\mathrm{flake}$ is calculated based on the optical transfer matrix scheme, previously described in Reference \cite{abtahiThicknessDependenceLinear2026} and $A$ is the slowly varying attenuation of the fiber network, which is parameterized as a spline with three support points with $y$-coordinates $y_i$. Optimizing the thickness $t$ and $y_i$, we obtain excellent agreement with the recorded transmission data, as displayed in Figure~\ref{fig:crystal_thickness_fit}(b), and we obtain a thickness value of 408.28 nm for the material.

\section{Multi-Process Photoluminescence Model}
As described in the main text, we observed substantial deviations of our recorded CAR values from the typical $\propto 1/P$ behavior, due to the nonlinear drop of the background PL signal at low powers, which is displayed in Figure~\ref{fig:single_countrates}. To explicitly model this, we fit the function
\begin{align}
	R(P) = \alpha + \beta P + \frac{\gamma P P_\mathrm{s}}{P + P_\mathrm{s}}
\end{align}
with the parameters $\alpha$, $\beta$, $\gamma$, and $P_\mathrm{s}$ to the count rates of each channel. Note that the product of $\gamma$ and $P_\mathrm{s}$ yields the saturated count rate $R_\mathrm{s}$. For both channels we obtain similar values for the saturation power $P_\mathrm{s} = (1.6\pm0.3)\,\mathrm{mW}$, and calculate corresponding saturation count rates of $R_\mathrm{s} = (5.6 \pm 1.2)\,\mathrm{kHz}$ and $R_\mathrm{s} = (3.6 \pm 0.7)\,\mathrm{kHz}$, respectively. Furthermore, we obtain a negative value of $\alpha = -(1.22 \pm 0.12)\,\mathrm{kHz}$ for both channels, which implies a cut-on effect for $R(P)$ around $P = 0.36\,\mathrm{mW}$. This could be highly interesting for boosting the CAR further; however, for verification, more data points at lower power are needed.

\begin{figure}[htbp]
	\centering
	\includegraphics[scale=0.7]{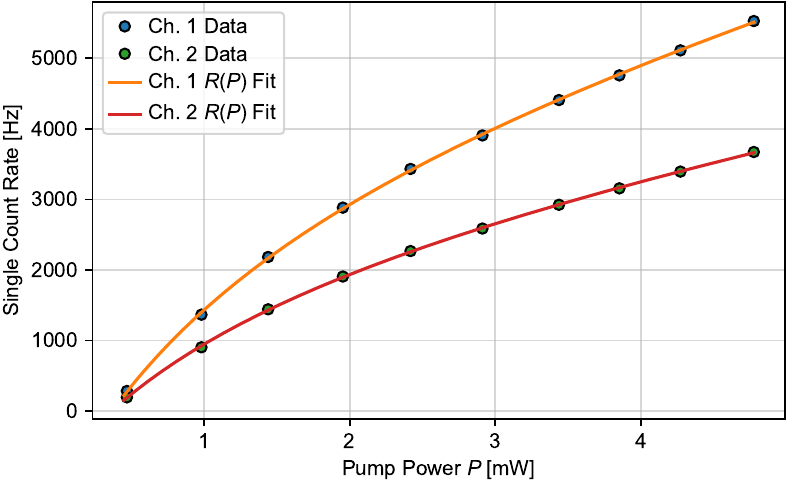}
	\caption{Single count rates as a function of pump power $P$ for each of the time-to-digital converter channels.}
	\label{fig:single_countrates}
\end{figure}

\section{Unitary Behavior of the 50:50 Splitter}
Prior to the coincidence measurement, we analyzed the fiber-based 50:50 splitter with respect to its compatibility with a unitary model, i.e. polarization insensitivity. For this, we used a second polarimetry setup of identical design to the one of the reference output, to characterize the light exiting the 780 nm single-mode fiber in the sample branch of the setup. By setting the polarization controller to random settings, we generated 100 random polarization state pairs for the sample and reference outputs.

In Figure~\ref{fig:ref_to_sample_state_mapping}(a), we have displayed histograms of the degree of polarization (DoP) of each recorded state. We find that, for the most part, the recorded states are highly polarized. For the reference output, we find that the DoP values are clustered nearly symmetrically around 1. This indicates that, in this measurement, we are likely not limited by the DoP of the light source, but the polarimetry device instead. A potential explanation for this could be deviations from a perfect quarter-wave retardance for the quarter-wave plate, which we assumed as ideal in our Stokes vector calculation. For the reference output we observe a slight decrease of the DoP, which could indicate that the splitter depolarizes the light to a small degree.

\begin{figure}[htbp]
	\centering
	\includegraphics{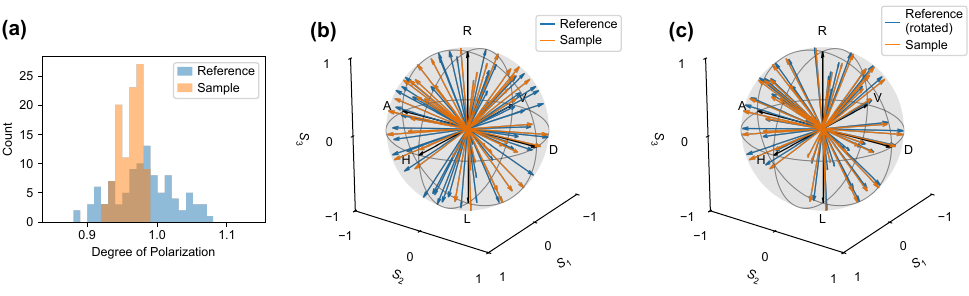}
	\caption{
		\textbf{Polarization state analysis of the 50:50 splitter outputs.}\\
		\textbf{(a)} Histograms of the degree of polarization for the recorded states of each output.
		\textbf{(b)} Measured polarization states for the reference and sample outputs.
		\textbf{(c)} Reference states mapped onto their corresponding sample states via a single, optimized rotation transform.
		The mapped reference states are in good agreement with the sample states, which indicates that the polarization states are connected via a unitary transformation.
		Note that in (b) and (c) only half of all recorded states are shown for improved legibility. 
	}
	\label{fig:ref_to_sample_state_mapping}
\end{figure}

Figure~\ref{fig:ref_to_sample_state_mapping}(b) displays the normalized Stokes parameters of the recorded polarization of the reference and sample outputs. Only half the states are shown for better readability. We observe that, while some systematic relation between the two outputs is present, most state vectors do not align perfectly. To test if a unitary transformation between the ensembles of states exists, we use Kabsch's algorithm \cite{kabschSolutionBestRotation1976} to find the optimal rotation for mapping the two sets of vectors onto each other. Mapping the polarization states of the reference output onto the sample output, we obtain excellent agreement, as shown in Figure~\ref{fig:ref_to_sample_state_mapping}(c).

\section{Co-Polarized Tomography Scheme} \label{sec:copol_tomo}
To show that the polarization state
\begin{align}
	\ket{\psi_\mathrm{G}} = \frac{\sin(\varphi_\mathrm{P})}{\sqrt{2}} \left( \ket{HH} - \ket{VV} \right)
	+\frac{e^{i\delta_\mathrm{P}} \cos(\varphi_\mathrm{P})}{\sqrt{2}} \left( \ket{HV} + \ket{VH} \right), \label{eq:psiG_suppl}
\end{align}
generated by the in-plane $\chi^{(2)}$ elements of 3R-MoS\textsubscript{2} can be uniquely identified in a co-polarized tomography scheme, we explicitly write the analyzer states $\ket{HH}$,
\begin{align}
	\ket{DD} &= \frac{1}{2} ( \ket{HH} + \ket{HV} + \ket{VH} + \ket{VV} ), \\
	\ket{RR} &= \frac{1}{2} ( \ket{HH} + i\ket{HV} + i\ket{VH} - \ket{VV} )
\end{align}
in the $\ket{H}$ and $\ket{V}$ polarization basis. The corresponding projection coefficients of $\ket{\psi_\mathrm{G}}$ are then:
\begin{align}
	\braket{HH}{\psi_\mathrm{G}} &= \frac{\sin(\varphi_\mathrm{P})}{\sqrt{2}} \\
	\braket{DD}{\psi_\mathrm{G}} &= \frac{e^{i\delta_\mathrm{P}} \cos(\varphi_\mathrm{P})}{\sqrt{2}} \\
	\braket{RR}{\psi_\mathrm{G}} &= \frac{1}{\sqrt{2}} \left( \sin(\varphi_\mathrm{P})  - i e^{i\delta_\mathrm{P}} \cos(\varphi_\mathrm{P})\right).
\end{align}
By calculating their absolute values squared, we obtain the coincidence rates for each projection measurement:
\begin{align}
	I_{HH} &= |\braket{HH}{\psi_\mathrm{G}}|^2 = \frac{I_0}{2} \sin^2(\varphi_\mathrm{P}) \label{eq:IHH}\\ 
	I_{DD} &= |\braket{DD}{\psi_\mathrm{G}}|^2 = \frac{I_0}{2} \cos^2(\varphi_\mathrm{P}) \label{eq:IDD}\\
	I_{RR} &= |\braket{RR}{\psi_\mathrm{G}}|^2 = \frac{I_0}{2} (1 + \sin(2\varphi_\mathrm{P}) \sin(\delta_\mathrm{P})). \label{eq:IRR}
\end{align}
Here, $I_0$ is the maximum detectable count rate. 
To simplify Equation~\eqref{eq:IRR}, we used
\begin{align}
	\left| \sin(\varphi_\mathrm{P}) - i e^{i\delta_\mathrm{P}} \cos(\varphi_\mathrm{P}) \right|^2
	&= \sin^2(\varphi_\mathrm{P}) + \cos^2(\varphi_\mathrm{P}) + 2\sin(\varphi_\mathrm{P})\cos(\varphi_\mathrm{P})\sin(\delta_\mathrm{P}) \notag \\ 
	&= 1 + \sin(2\varphi_\mathrm{P})\sin(\delta_\mathrm{P}).
\end{align}
From Equations \eqref{eq:IHH}-\eqref{eq:IRR}, it is straightforward to retrieve the parameters $I_0$, $\varphi_\mathrm{P}$, and $\delta_\mathrm{P}$:
\begin{align}
	I_0 &= I_{HH} + I_{DD} \\
	\varphi_\mathrm{P} &= \arctan(\sqrt{I_{HH} / I_{DD}}) \\
	\delta_\mathrm{P} &= \arcsin \left( \frac{2 I_{RR} - I_0}{I_0 \sin(2 \varphi_\mathrm{P})} \right). \label{eq:deltaP}
\end{align}
Note that, in the special case of $\varphi_\mathrm{P} = 0$, Equation~\eqref{eq:deltaP} becomes singular, as $\delta_\mathrm{P}$ has no influence on the generated state. This scheme works analogously when using the orthogonal counterparts of the analyzer states. If a known unitary transformation $U \otimes U$ is applied to the generated state, the analyzer states can be adjusted accordingly, e.g. $\ket{H'} = U \ket{H}$.

\section{Basis Choice for Linear Pump States} \label{sec:basis_choice}
In the case of linear pump polarization $\delta_\mathrm{P} = 0$, we can always tailor the choice of our logical qubit basis states $\ket{0}$ and $\ket{1}$ depending on $\varphi_\mathrm{P}$ such that $\ket{\psi_\mathrm{G}}$ corresponds to the maximally entangled Bell state $\ket{\psi^+} = \frac{1}{\sqrt2} (\ket{01} + \ket{10})$. Specifically:
\begin{align}
\begin{split}
	\ket{0} &= \cos(\varphi_\mathrm{P} / 2) \ket{H} - \sin(\varphi_\mathrm{P} / 2) \ket{V} \\
	\ket{1} &= \sin(\varphi_\mathrm{P}/ 2) \ket{H} + \cos(\varphi_\mathrm{P} / 2) \ket{V},
\end{split} \label{eq:basis_choice}
\end{align}
which corresponds to a rotation of the coordinate system around the direction of light propagation by $\varphi_\mathrm{P}/2$. Note that this is also a unitary transformation and therefore offers the degree of freedom observed numerically in searching for an optimal pair of unitary transformations.
Equations \eqref{eq:basis_choice} can be verified by explicitly calculating $\ket{\psi^+}$. For this, we first calculate the individual components
\begin{align}
	\ket{01} = \ket{0}\otimes\ket{1}
	&= \cos(\varphi_\mathrm{P} / 2) \sin(\varphi_\mathrm{P} / 2) \ket{HH} \notag\\
	&\quad+ \cos^2(\varphi_\mathrm{P} / 2) \ket{HV} \notag\\
	&\quad- \sin^2(\varphi_\mathrm{P} / 2) \ket{VH} \notag\\
	&\quad+ \cos(\varphi_\mathrm{P} / 2) \sin(\varphi_\mathrm{P} / 2) \ket{VV} \\
	\ket{10} = \ket{1}\otimes\ket{0}
	&= \cos(\varphi_\mathrm{P} / 2) \sin(\varphi_\mathrm{P} / 2) \ket{HH} \notag\\
	&\quad- \sin^2(\varphi_\mathrm{P} / 2) \ket{HV} \notag\\
	&\quad+ \cos^2(\varphi_\mathrm{P} / 2) \ket{VH}  \notag\\
	&\quad+ \cos(\varphi_\mathrm{P} / 2) \sin(\varphi_\mathrm{P} / 2) \ket{VV}
\end{align}
which add up to
\begin{align}
	\ket{\psi^+} = \frac{1}{\sqrt{2}} \left( \sin(\varphi_\mathrm{P}) (\ket{HH} - \ket{VV}) + \cos(\varphi_\mathrm{P}) (\ket{HV} + \ket{VH}) \right),
\end{align}
where we used the trigonometric identities $2\cos(x)\sin(x) = \sin(2x)$ and $\cos^2(x)-\sin^2(x) = \cos(2x)$.
This is $\ket{\psi_\mathrm{G}}$ for the case of $\delta_\mathrm{P} = 0$.

\end{adjustwidth}

\end{document}